\documentclass[12pt]{article}
\usepackage{graphicx}
\usepackage{amsfonts}

\begin{document}
\begin{titlepage}

\hfill{Preprint {\bf SB/F/03-311}} \hrule \vskip 2.5cm

\centerline{\bf Phase space interference and the WKB approximation
for squeezed number states} \vskip 2cm \centerline{D.F. Mundarain
and J. Stephany} \vskip 4mm

\centerline{\it  Departamento de F\'{\i}sica, Universidad Sim\'on Bol\'{\i}var,}
\centerline{\it Apartado 89000, Caracas 1080A, Venezuela.}

\begin{abstract}
Squeezed number states for a single mode Hamiltonian are
investigated from two complementary points of view. Firstly the
more relevant features of their photon distribution are discussed
using the WKB wave functions. In particular the oscillations of
the distribution and the parity behavior are derived and compared
with the exact results. The accuracy is verified and it is shown
that for high photon number  it fails to reproduce the true
distribution. This is contrasted with the fact that for moderate
squeezing the WKB approximation gives the analytical justification
to the interpretation of the oscillations as the result of the
interference of areas with definite phases in phase space. It is
shown with a computation at high squeezing using a modified
prescription for the phase space representation which is based on
Wigner-Cohen distributions that  the failure of the WKB
approximation does not invalidate the area overlap picture.

\end{abstract}\vskip 2cm
\hrule
\bigskip
\centerline{\bf UNIVERSIDAD SIMON BOLIVAR}
\vfill
\end{titlepage}

\section{Introduction}
The investigation in phase space of the properties of quantum
systems contributes with a remarkable picture of those systems in
which the states are described by areas in such a way that the
emergence of diverse quantum phenomena
\cite{Wheeler,Schleich,Walls} appears as the result of the
interference  of the portions of these areas which overlap. The
area which represent each state may be fixed from
considerations in the context of the old Bohr-Sommerfeld approach
\cite{Wheeler, Schleich}, with the aid of phase-space
distributions \cite{Walls} or  using the WKB approximate wave
functions \cite{Dowling}. This last approach has the advantage of
being more systematic but is constrained by the limitations of the
approximation. In this paper we illustrate both terms in this
sentence  discussing the photon distribution of squeezed number
states\cite{Yuen}. The photon distribution of squeezed number
states is parity sensible and present characteristic oscillations
with increasing photon number until it decay and vanish for
sufficiently large photon number. These oscillations which may be
computed exactly \cite{Yuen,Kim,Albano} are similar to the ones
which appear for squeezed states which  been discussed thoroughly
using the phase space description \cite{Schleich}. It is then
interesting to investigate also how this approach  may also be
applied to the case of  the squeezed number states. A first step
in this direction was taken in Ref. \cite{Kim} where it was shown
that the main characteristics of the distribution are consistent
with interference in phase space picture. Here we develop with
detail the WKB approach of the phase space picture to investigate
how the parity restrictions may be enforced and under which
conditions the approximation breaks down. The outcome is that the
WKB approximation is not able to represent the last maximum of the
distribution  in the the high photon number region. This failure
may be traced to to the limitations of the WKB approximation and
not to the phase space picture which as we show  below is still
realized (at least for high squeezing) using the more direct scheme to
assign the areas to the states which relies on the Wigner-Cohen
distributions.

\section{The WKB approximation}
\label{sec1}

Let us begin discussing the WKB approximation for squeezed number
states and its relation with the phase  space picture. We work
with the normalized one mode  Hamiltonian
\begin{equation}
 H = \frac{1}{2} p^2 + \frac{1}{2}x^2=a^{\dagger} a + \frac{1}{2}
\end{equation}
and denote by  $|n>$ the  eigenstates of $H$ and of the number
operator $N = a^{\dagger} a$ with eigenvalues  $n+ 1/2$ and  $n$.

The squeezed number states \cite{Yuen} are defined by
\begin{eqnarray}
|n,r\rangle = S(r) |n\rangle
\end{eqnarray}
with the squeeze operator  $S(r)$ given by
\begin{equation} \label{eq:SQUEOP}
S(r) = \exp\left( \frac{ 1}{2}r(a^{\dagger})^2 - \frac{1}{2} ra^2\right).
\end{equation}
(We are taking the phase parameter of the squeezing equal to zero
since the photon distribution does not depends on it). These
states are eigenstates with eigenvalues $(n+1/2)$ of the
transformed Hamiltonian $H^{\prime}=S(r)HS(r)^{-1}$ which is shown
to be equal to \cite{Yuen},
\begin{equation}
 H^{\prime} = \frac{1}{2} p^2 e^{-2 r} + \frac{1}{2}x^2 e^{2r} \ \ .
\end{equation}
Alternatively the squeezed number states may be interpreted as the
Fock states of a related system with mass $e^{2r}$ and the same
frequency. For a more detailed description of these states see
\cite{Yuen,Kim,Albano}.  The WKB wave functions for these states
are given by \cite{Merzbacher,Dowling},
\begin{equation}
\Phi_n^{(r)}(x) = \frac{2}{C_n}({T}_n^{(r)} \,p_n^{(r)}(x))^{-1/2}\,  \cos
\left( S_n^{(r)}(x) -\pi /4 \right),
\end{equation}
where $C_n$ is a normalization constant. The value of  $p_n^{(r)}$ is obtained from the equation
\begin{equation}
n+1/2 = \frac{1}{2} (p_n^{(r)}(x))^2 e^{-2 r} + \frac{1}{2}x^2 e^{2 r}
\end{equation}
and is given in terms of the classical turning point value $\varepsilon_n^{(r)}=
e^{-r}\sqrt{2m+1}$ by
\begin{eqnarray}
p_n^{(r)}(x) =& e^{2r} ((\varepsilon_n^{(r)})^2-x^{2})^{1/2}\nonumber\\
 =& e^{ r} p_n^{(0)}(x e^{r})
\end{eqnarray}
The phase function is given by,
\begin{eqnarray}
S_n^{(r)}(x) =& \int_x^{\varepsilon_n^{(r)}} p_n^{(r)}(x^{\prime}) dx^{\prime }
\nonumber\\=&S_n^{(0)}(x e^{r}) \ \ ,
\end{eqnarray}
and the normalization constant has been written in terms of the period of the related
classical motion,
\begin{eqnarray}
T_n^{(r)}=& \int_{-\varepsilon_n^{(r)}}^{\varepsilon_n^{(r)}}
 \frac{ dx^{\prime }}{p_n^{(r)}(x^{\prime})}=&2\pi \, e^{-2r}
\end{eqnarray}

For $r=0$ we recover the description of the number states. The WKB wave
functions are compressed by the squeezing in the
same form as the exact states,
\begin{equation}
\Phi_n^{(r)}(x) = e^{r/2} \Phi_n^{(0)}(e^r x) \ .
\end{equation}

We note that $p_n^{(r)}$ is a symmetric function and that
\begin{equation}
\int_{-\varepsilon_n^{(r)}}^{\varepsilon_n^{(r)}} p_n^{(r)}(x^{\prime})
dx^{\prime}  = \pi/2  + n \pi \ \ \ .
\end{equation}

Hence the phase function $S_n(x)$ satisfies the relation
 \begin{equation}\label{pode1}
S_n^{(r)}(-x) -\pi/4 = -\left( S_n(x) -\pi/4 \right) + n \pi
\end{equation}
and one can show that the WKB wave functions  satisfy the parity conditions of the
exact states.
\begin{eqnarray}
\Phi_n^{(r)}(-x)& =& (-1)^n   \Phi_n^{(r)}(x)
\end{eqnarray}
Again for $r=0$ we are dealing with the number states.

The WKB wave functions are meaningless near the classical return points where they
diverge. Also we note that
the approximation holds better for higher values of $n$. The
normalization of the wave functions is such that
\begin{eqnarray}
|C_n|^2=1+\frac{1}{\pi}
\int_{-\varepsilon_n^{(r)}}^{\varepsilon_n^{(r)}}\frac{\sin(2S_m^{(r)}(x))}
{p_m^{(r)}(x)}dx
\end{eqnarray}
with the  integral in the right term  vanishing  for large n.

\section{Photon statistics in the WKB approximation}\label{sec3}

\begin{figure}\label{WKB2}
\includegraphics[scale=0.4]{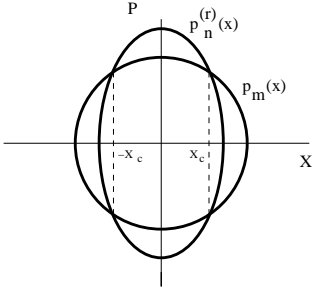}
\caption{The four points in phase space which contribute to the photon distribution amplitude}
\end{figure}
Now we turn to the computation of the photon distribution of the squeezed number
states within this  approximation. It is given by
\begin{equation}
P_{m n} = | W_{m n}|^2
\end{equation}
with
\begin{equation}
 W_{m n} = <m|n,r> = \int_{-\infty}^{\infty}\Phi_m^{(0)}(x)\Phi_n^{(r)}(x) dx .
\end{equation}
Using the WKB wave functions is shown that, as was suggested in References \cite{Dowling, Kim}
the probability amplitude   $W_{m n}$ may be  expressed in terms of a phase $\varphi_{m n}$ and an overlapping area $A_{m n}$
in the form,
\begin{eqnarray}\label{Ec2}
W_{m n}& =& (A_{m n})^{ 1/2} e^{i \varphi_{m n}} +(A_{m n})^{ 1/2} e^{-i \varphi_{m n}}\nonumber\\
& +& (A_{m n})^{ 1/2} e^{i \varphi_{m n}^{\prime}} +(A_{m n})^{ 1/2} e^{-i \varphi_{m n}^{\prime}}
\end{eqnarray}

In this expression $A_{mn}$ is given by
\begin{equation}
A_{m n} = 2 \pi \, \left( T_m \, T_n^{(r)} \, p_m^2(X_c)\right)^{-1}  \,|\frac{d}{dx} p_m (X_c)-\frac{d}{dx} p_n^{(r)} (X_c)|^{-1}
\end{equation}
and is shown to be the overlapping area between a circular ring of interior radius
$\sqrt{2m}$ and exterior radius $\sqrt{2m+2}$ of total area $2\pi$ representing the $m$ Fock state in phase space
and a deformed elliptical ring of equal area which corresponds to the $n$ squeezed number state.
The phases are given by
\begin{eqnarray}
\varphi_{m n } =& S_m^{(0)} (X_c) - S_n^{(r)} (X_c) - \pi/4
\ , \nonumber \\
\varphi_{m n}^{\prime} =& S_m^{(0)} (-X_c) - S_n^{(r)} (-X_c) + \pi/4
\end{eqnarray}
where $X_c$ is the point where,
\begin{equation}
p_m^{(0)}(X_c) = p_n^{(r)} (X_c)
\end{equation}
as can be shown in Figure (\ref{WKB2}).
We have then,
\begin{equation}
X_c =\left( \frac{e^{2r} (2 n +1) -( 2 m +1)}{(e^{4r} -1)}\right)^{1/2} .
\end{equation}
From the properties of $S_m^{(r)}$ (Ec. \ref{pode1}) we obtain:
\begin{eqnarray}\label{Ec1}
\varphi_{m n}^{\prime}&= & -\varphi_{m n} + (m-n) \pi
\end{eqnarray}
Using (\ref{Ec1}) and (\ref{Ec2}) the probability amplitude is given by
\begin{equation}
W_{m n} =   2 (A_{m n})^{1/2} \cos (\varphi_{m n}) \left( 1+\cos \left( (m-n)\pi\right) \right).
\end{equation}
This result is in agreement with the form of the amplitude proposed  in  \cite{Kim} and
in particular  \ref{Ec1} gives the correct parity behavior of the distribution.
\begin{figure}\label{dfm3}
\centerline{\includegraphics[scale=0.5]{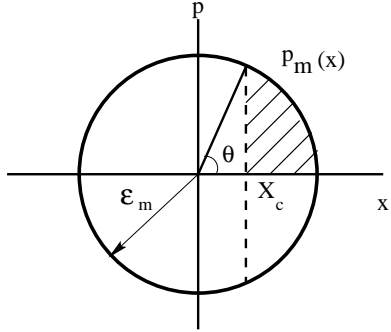}}
\caption{Geometrical interpretation of $S_m(X_c)$}
\end{figure}

The connection with the geometrical construction is obtained
 noting that $S_m^{(0)}(X_c)$
is given by the shadowed area in Fig.(\ref{dfm3}).
Then,\begin{equation}
S_m^{(0)}(X_c) = \frac{1}{2} \theta (\varepsilon_m^{(0)})^2 - \frac{1}{2} X_c
\left((\varepsilon_m^{(0)})^2-X_c^2\right)^{1/2}
\end{equation}
where
\begin{equation}
\theta = \arccos \{ \frac{X_c}{\varepsilon_m^{(0)}} \}
\end{equation}
In the same way we have,
\begin{equation}
S_n^{(r)}(X_c) = \frac{1}{2} \theta (\varepsilon_n^{(r)})^2 - \frac{1}{2} X_c
\left((\varepsilon_n^{(r)})^2-X_c^2\right)^{1/2}
\end{equation}
where
\begin{equation}
\theta = \arccos \{ \frac{X_c}{\varepsilon_n^{(r)}} \}
\end{equation}

\vspace{1.0cm}
\begin{figure}[t]
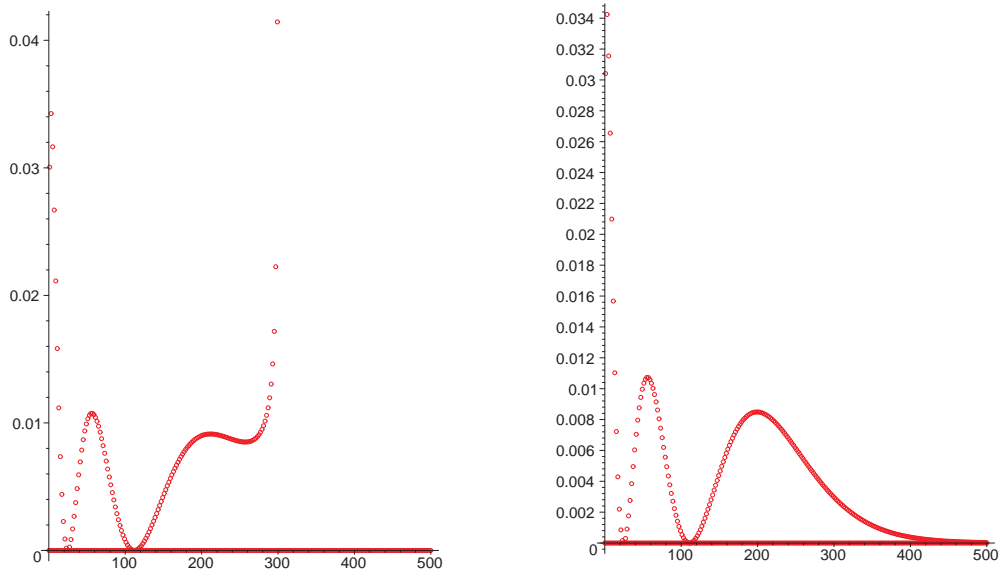

\includegraphics[scale=0.4]{WKBPm52.ps}\includegraphics[scale=0.4]{EXPm52.ps}
\caption{Photon distribution $P_{m n}$ for a squeezed number state with $n=5$
and $r=2$. Left: WKB approximation. Right:Exact computation}\label{WKBPm52}
\end{figure}

In Fig (\ref{WKBPm52}) we show the  WKB approximation for the photon distribution of a
squeezed number state with $n=5$ and $r=2$ compared with the exact computation.
The approximation is faithful for photon numbers
below the last maximum.

 \section{Wigner-Cohen distributions and interference in phase space}
\label{sec4}

The failure of the approximation presented in the last section for large
photon numbers is related to the the breakdown of the validity of the WKB
approximation. Nevertheless as we show below it does not exclude the description of this
part of the distribution using the area overlapping in phase space. To see this let us
introduce $ F_n^{(r)}(x,p)$ to denote any of the phase space Wigner-Cohen distributions
\cite{Cohen}  associated to the squeezed number states which satisfies,

\begin{equation}
\int F_n^{(r)}(x,p) dp = |\psi_n^{(r)} (p)|^2
\end{equation}
and
\begin{equation}
\int F_n^{(r)}(x,p) dx = |\overline{\psi}\,_n^{(r)} (p)|^2
\end{equation}
where $\psi_n^{(r)} (x)$ y $ \overline{\psi}\,_n^{(r)} (p)$
are the wave functions of the squeezed number states in the position
and momentum representations respectively.

The photon number distribution may be obtained by integrating $ F_n^{(r)}(x,p)$
over the $m$-th ring $\Omega_m$  associated to the number state $|m>$
which is the area between to circles of radii $\sqrt{2 m}$
and $\sqrt{2 m +2}$ respectively,
 \begin{equation}
A_{m n} = \frac{1}{2}\quad \int_{\Omega_m} dpdx  F_n^{(r)}(x,p)
\end{equation}
For a highly squeezed number state the integral in
the configuration space variable $q$ can
approximately taken over the whole line
\begin{equation}
A_{m n} \approx \int_{\sqrt{2 m}}^{\sqrt{2 m+2}} |\overline{\psi}_n^{(r)} (p)|^2 dp
\end{equation}
and by supposing a smooth behavior in the ring we get,
\begin{equation}
A_{m n} \approx |\overline{\psi}\,_n^{(r)} (\sqrt{2 m+1}) |^2 (\sqrt{2 m+2}-\sqrt{2 m} )
\end{equation}
Taking into account the parity of the functions we finally have,
\begin{equation}
P_{m n} \approx  \left[1 + \cos((m-n)\pi)\right]^2 A_{m n}
\end{equation}
Substituting the values of the waves functions we obtain,
\begin{eqnarray}\label{Ec22}
P_{m n}&\approx& \left(1 + \cos((m-n)\pi)\right)^2\left( \sqrt{2 m+2} -  \sqrt{2 m} \right) \nonumber\\
&&e^{-r}\frac{ e^{- (2 m+1) e^{-2r}} (H_n(\sqrt{2 m +1} e^{-r}))^2}{\sqrt{\pi} 2^n  n! } \end{eqnarray}
\begin{figure}[t]
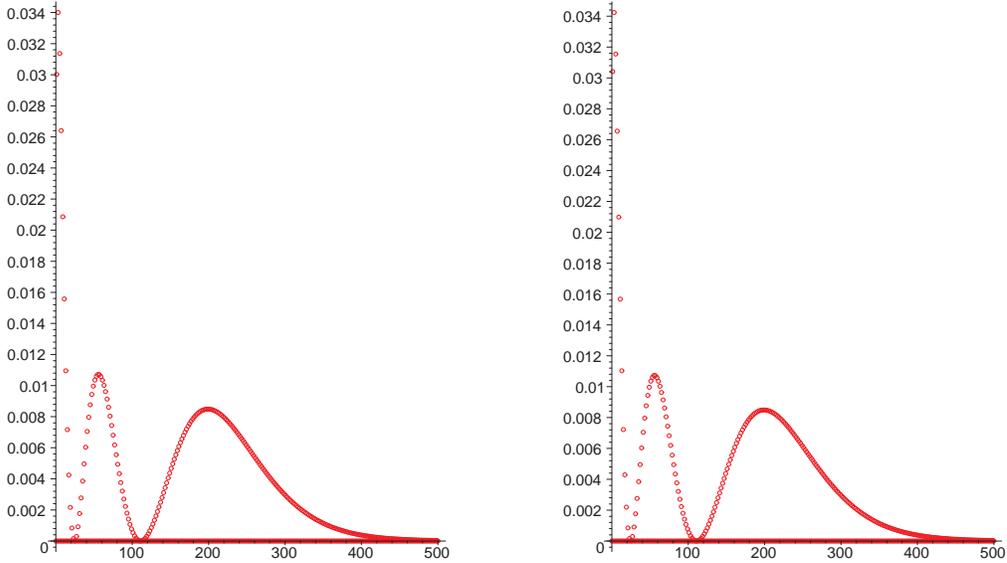

\includegraphics[scale=0.4]{Pm52.ps}\includegraphics[scale=0.4]{EXPm52.ps}
\caption{Photon distribution  $P_{m n}$ for a squeezed number state $n=5$ and $r=2$.
Left: Phase space computation(\ref{Ec22}). Right: Exact result}\label{Pm52}
\end{figure}
which is in an excellent agreement with the exact value as shown in
Fig. (\ref{Pm52}) in the particular case of $n=5$ and  $r=2$.
The agreement stills holds for photon numbers beyond the last oscillation
where the WKB approximation breaks down. This expression also shows that for
large squeezing the photon number oscillation are driven by the oscillations of
the wave function in momentum space.

\section{Conclusion}
We have computed the oscillations  photon number
distribution of squeezed number states within
the WKB  approximation making connection with the phase space description given in
Ref.\cite{Dowling}. In this case there are four sectors where the areas
overlap and give a contribution to the probability amplitude \cite{Kim}.
We show that for high photon number the WKB approximation fails to
describe the last oscillation.
Finally we note that a description in terms of interference of the overlapping areas
in phase space may be recovered, but not  based in the WKB approximation but in
the Wigner-Cohen distributions. For large squeezing this prescription reproduces
the photon number distribution for the whole range.


\begin{thebibliography}{99}
\bibitem{Wheeler} J. A. Wheeler, Lett. Math. Phys.  {\bf 10}, 201 (1985).
\bibitem{Schleich} W. Schleich y J.A. Wheeler, {\it Nature} 326, 574 (1987); W. Schleich,
H. Walther and J. A. Wheeler, Foundations of Physics{\bf 18}, 953 (1988).
\bibitem{Walls} W. Schleich, D. F. Walls and J. A. Wheeler, Phys. Rev. A
{\bf 38}, 1177 (1988)
\bibitem{Dowling} J. P. Dowling, W.P. Schleich y J.A. Wheeler, {\it Ann. der Phys.} {\bf 7},
423 (1991).
\bibitem{Yuen} H. P. Yuen, {\it  Phys. Rev.} A {\bf 13}, 2226 (1976).
\bibitem{Kim} M . S. Kim, F.A.M. de Oliveira y P.L. Knight, {\it Phys. Rev.} A {\bf 40}, 2494  ( 1989 ); M. S. Kim, F. A. M. De Oliveira and P. L. Knight, Optics Comm.
{\bf 72}, 99 (1989).
\bibitem{Albano} L.Albano, D.F.Mundarain and J.Stephany,{\it  J.Opt.B:Quantum Semiclass.Opt.}
{\bf 4}, {352-357 (2002)}.
\bibitem{Merzbacher} E.Merzbacher, {\it Quantum Mechanics}, 2nd Ed., (Wiley, New York, 1970)
\bibitem{Cohen} L. Cohen, {\it Frontiers of Non equilibrium Statistical Physics}
(eds. G. T. Mooore $\&$ M. O. Scully) 97 ( Plenum, New York, 1986).
\end{thebibliography}
\end{document}